\documentclass[english,american]{article}
\usepackage[latin9]{inputenc}
\usepackage{geometry}
\usepackage{amsthm}
\usepackage{amsmath}
\usepackage{amssymb}
\usepackage{graphicx}
\usepackage{esint}

\makeatletter

\newcommand{\lyxdot}{.}

\theoremstyle{plain}
\newtheorem{thm}{\protect\theoremname}
\theoremstyle{definition}
\newtheorem{defn}[thm]{\protect\definitionname}
\theoremstyle{remark}
\newtheorem{rem}[thm]{\protect\remarkname}

\usepackage{amsfonts}
\usepackage{epsfig}\usepackage[american]{babel}
\usepackage{graphicx}

\providecommand{\remarkname}{Remark}
\providecommand{\theoremname}{Theorem}

\providecommand{\remarkname}{Remark}
\providecommand{\theoremname}{Theorem}

\makeatother

\usepackage{babel}
\addto\captionsamerican{\renewcommand{\definitionname}{Definition}}
\addto\captionsamerican{\renewcommand{\remarkname}{Remark}}
\addto\captionsamerican{\renewcommand{\theoremname}{Theorem}}
\addto\captionsenglish{\renewcommand{\definitionname}{Definition}}
\addto\captionsenglish{\renewcommand{\remarkname}{Remark}}
\addto\captionsenglish{\renewcommand{\theoremname}{Theorem}}
\providecommand{\definitionname}{Definition}
\providecommand{\remarkname}{Remark}
\providecommand{\theoremname}{Theorem}

\begin{document}

\title{A Fractional Micro-Macro Model for Crowds of Pedestrians based on
Fractional Mean Field Games}

\author{Ke-cai~Cao%
\thanks{caokc@njupt.edu.cn%
},\\
College of Automation, Nanjing University of Posts and Telecommunications,
\\
Nanjing, P.R. China, 210023.\\
YangQuan Chen %
\thanks{ychen53@ucmerced.edu%
},\\
Mechatronics Embedded Systems and Automation Lab, School of Engineering,
\\
University of California, Merced, CA, USA 95343.\\
Dan Stuart %
\thanks{idahoeinstein@gmail.com%
},\\
Department of Electrical and Computer Engineering, \\
Utah State University, Logan, UT, USA 84322.}
\maketitle
\begin{abstract}
Modeling of crowds of pedestrians has been considered in this paper
from different aspects. Based on fractional microscopic model that
may be much more close to reality, a fractional macroscopic model
has been proposed using conservation law of mass. Then in order to
characterize the competitive and cooperative interactions among pedestrians,
fractional mean field games are utilized in the modeling problem when
the number of pedestrians goes to infinity and fractional dynamic
model composed of fractional backward and fractional forward equations
are constructed in macro scale. Fractional micro-macro model for crowds
of pedestrians are obtained in the end. Simulation results are also
included to illustrate the proposed fractional microscopic model and
fractional macroscopic model respectively.

\noindent {\em Keywords:} Fractional Mean Field Games, Microscopic
Model, Macroscopic Model, Micro-Macro Model, Fractional Calculus.
\end{abstract}

\section{Introduction }

Methodology for modeling of crowds of pedestrians has been categorized
as micro scale, macro scale and meso scale in previous research. It
is reasonable to choose different models in different scenarios as
``All models are wrong but some of them are useful '' (quote George
E. P. Box) in \cite{BruceJ.West2014}. Thus, no models are perfect
for all scenarios.

\subsection{Short review of modeling for crowds of pedestrians}

A lot of work has been done for microscopic model since Dirk Helbing's
work of \cite{HelbingDirk1995,Helbing2000} because the framework
of social forces are similar to the framework of Newton's principle
and it is not difficult to understand. Another reason for the widespread
use of this social force model lies in that heterogeneity of each
pedestrian such as mobilities or reactions can be considered explicitly.
Thus not only theoretical work but also simulation results have gained
a lot of attention such as \cite{Bellomo2011a}, \cite{Couzin2005},
\cite{Couzin2008}, \cite{WeiguoSong2006} and \cite{NirajanShiwakoti2011}.
One thing should be pointed out is that the burden of computation
in micro scale has imposed great challenges when the number of pedestrians
goes to infinity and some elements such as pedestrian's memory, long
range interactions or other statistical characters have been seldom
considered in previous work. The disadvantages of computation burden
in microscopic model have been successfully removed in macroscopic
model as all pedestrians are treated as uniform physical particles.
Thus different kinds of macroscopic models have been published based
on the conservation law of mass and momentum such as \cite{Kachroo2009},\cite{Helbing1998c},\cite{RogerL.2002},\cite{Hughes2003},
high order macroscopic model in \cite{Jiang2010a}, nonlinear macroscopic
model in \cite{Sadeq2006} and coupled macroscopic-microscopic model
in\cite{CorradoLattanzio2011}. Although the computational burden
in macroscopic model has been reduced greatly compared with that in
microscopic model, main disadvantages of macroscopic model are that
individual characters of each pedestrian have been ignored and heterogeneity
of different pedestrians can not be characterized in the macro scale.

The authors believe that there are something important that have been
neglected in previous research and their effects should be included
in the problem of modeling and control of crowds so that obtained
results are close to reality.
\begin{enumerate}
\item Fractal time should be considered;

Movement of human beings are results of complex interactions from
physical part, psychological part and some reasons that are hard to
explain now. Inter-event time has been proved to be an important role
in characterizing people's movement as shown in \cite{BruceJ.West2014}.
The fact is that the distribution of inter-event time in our real
life satisfy one form of power law in most cases while distribution
of exponential form has been always assumed in previous research using
calculus of integer order. Thus fractional order of time scale should
be considered in characterizing movement and decision process of human
beings;

\item Fractal space should be considered;

Another important thing should be pointed out is that in previous
research, the time scale of each pedestrian is assumed to be uniform
and the dimensions of space are restricted to 1D, 2D and 3D. But these
assumptions are only reasonable if the crowds of pedestrians can fill
space like particles of gases or fluids while it is not the case in
most of the cases. Thus only normal diffusive process have been considered
in previous research and there are few results have been conducted
under sub-diffusive process or super-diffusive process that characterized
by fractal space;

\item Long range interactions have been considered in the schooling of fish,
flocking of birds and control of multi-agent systems and effects of
long range interactions that dominating system's phase transition
have just received a lot of attention recently. Based on obtained
results in \cite{Ishiwata2012}, we can say that long range interactions
in micro scale are connected with the fractional dynamics in macro
scale.
\end{enumerate}

\subsection{Modeling and Control based on Mean Field }

For crowds of pedestrians with large numbers, it is impossible and
not necessary to consider all the interactions one by one. In previous
research, methods based on mean field have been proposed to approximate
the mass effects of these interactions for physical system, financial
system and social dynamic system and the readers are referred to \cite{Achdou2012}
\cite{Achdou2012,Caines2009,Olivier2009,Dogbe2010}. Basic idea of
mean field framework is replacing all the interactions with an average
interaction in ``mean-field'' form to relieve the burden of computation
on each agent.

Mean field theory has been applied to control of multi-agent systems
in \cite{Nourian2011,Nourian2012a,Nourian2013} where decentralized
consensus protocols and decentralized optimizing algorithms are considered
using the philosophy of mean field. Mean field theory has also been
applied to the modeling problem for crowds of pedestrians in recent
years. For example, coupled dynamic model composed of backward Hamilton-Jacobi-Bellman
equations and forward Fokker-Planck equations have been presented
using mean-field limit approach in \cite{Dogbe2010}; Phenomenons
that occurring in two-population's interactions such as congestion
and aversion have been modeled using the method of mean field games
in \cite{Lachapelle2011} where coupled dynamic model composed of
backward Hamilton-Jacobi-Bellman equations and forward Fokker-Planck
equations are obtained; The mean field games theory has also been
used to construct traffic model in macro scale based on interactions
in micro scale in \cite{GeoffroyChevalier2015} while fractional dynamic
games has been used in \cite{Bogdan2011a} to construct dynamic models
for crowds of pedestrians.

With the help of calculus of fractional order, the authors of this
paper try to include the fractal time, fractal space and statistical
characters that have been neglected in previous research in the modeling
of crowds so that obtained models could be much more close to reality.
Based on our previous work on fractional modeling of crowds \cite{Ke-CaiCao2012,KecaiCao2015,Ke-CaiCao2015},
fractional mean field games theory has been investigated in this paper
to describe the competitive and cooperative interactions among pedestrians.
The rest of the paper is organized as follows. Fractional microscopic
model, fractional macroscopic model and fractional dynamic model based
on mean-field games are presented in Section \ref{sec:Main-Results}.
Simulation results for the proposed fractional macroscopic model and
fractional microscopic model have been shown in Section \ref{sec:Simulation-Results}.

\section{Preliminaries}

The following definitions of fractal derivative and Lemmas that will
be used in the followings are firstly presented for the easy of reading.
\begin{defn}
\cite{Parvate2005} \label{def:1} For a set $F\subset R$ and a subdivision
$P_{[a,b]}$, $a<b$, the mass function $\gamma^{\alpha}(F,a,b,)$
is given by
\[
\gamma^{\alpha}(F,a,b)\!\!=\!\!\lim_{\delta\rightarrow0}\inf_{\{P_{[a,b]}:|P|<\delta\}}\!\!\sum_{i=0}^{n-1}\!\frac{(x_{i+1}\!-\! x_{i})^{\alpha}}{\Gamma(\alpha+1)}\theta(F,\![x_{i},x_{i+1}]),
\]
where $\theta(F,[x_{i},x_{i+1}]=1$ if $F\cap[x_{i},x_{i+1}]$ is
non-empty, and zero otherwise, $P_{[a,b]}$ is a subdivision of the
interval $[a,b]$ and
\[
|P|=\max_{0\leq i\leq n-1}(x_{i+1}-x_{i}),
\]
the infimum being taken over all subdivisions P of $[a,b]$ such that
$|P|<\delta$.
\end{defn}

\begin{defn}
\cite{Parvate2005}\label{def:2} Let $a_{0}$ be an arbitrary but
fixed real number. The integral staircase function $S_{F}^{\alpha}(x)$
of order $\alpha$ for a set $F$ is given by
\[
S_{F}^{\alpha}(x)=\left\{ \begin{array}{cc}
\gamma^{\alpha}(F,a_{0},x) & \quad if\quad x\geq a_{0}\\
-\gamma^{\alpha}(F,a,x_{0}) & \quad otherwise.
\end{array}\right.
\]

\end{defn}

\begin{defn}
\cite{Parvate2005} \label{defn} The fractal derivative for $F^{\alpha}-$derivative
of $f$ at $x$ is
\begin{equation}
\mathcal{D}_{F}^{\alpha}(f(x))=\mbox{F-}\lim_{y\rightarrow x}\frac{f(y)-f(x)}{S_{F}^{\alpha}(y)-S_{F}^{\alpha}(x)}
\end{equation}
if the limit exists.

From definition \ref{def:1} to definition \ref{defn} listed above,
it is easy to see that the definition of integer order can be treated
as one special case of fractal derivative when $\alpha=$1. Thus the
fractal calculus offers us much more freedom in modeling dynamics
behaviors where ordinary differential equations and methods of calculus
of integer order are inadequate.
\end{defn}

\section{Main Results\label{sec:Main-Results}}

\subsection{Fractional Microscopic Model}

\begin{figure}[tbh]
\begin{centering}
\includegraphics{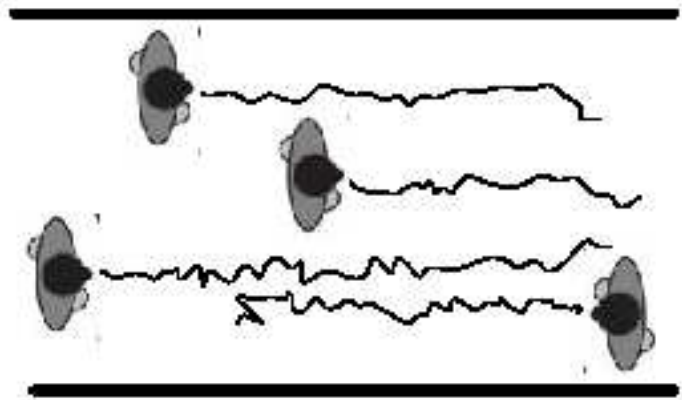}
\par\end{centering}

\protect\caption{Zig-Zag phenomenon in movement of each pedestrian\foreignlanguage{english}{ }}

\label{zigzag}
\end{figure}

The following dynamic model of integer order has been extensively
used in previous research of particles, human beings or some other
agents in micro scale
\begin{equation}
\left\{ \begin{alignedat}{1}\frac{dx_{i}}{dt} & =v_{i},\\
m_{i}\frac{dv_{i}}{dt} & =f_{i}^{S}+\sum_{j=1}^{n}f_{ij}^{N}+\sum f_{k}^{W},
\end{alignedat}
\right.\label{eq:micro-integer}
\end{equation}
where $x_{i}$ is the position and $v_{i}$ is the velocity. one common
assumption has been made that movement of each pedestrian is continuous
and differentiable everywhere, That is the case if we observe the
movement of each pedestrian with a very large scale such as in macro
scale. However the condition of differentiable everywhere is hard
to be satisfied in reality. So will the $dx/dt$ give the true picture
of pedestrian's movement in micro scale or will the $d^{\alpha}x/dt^{\alpha}$
be much closer to reality when only continuous condition is satisfied.
Related research on this fractional aspect has been shown in \cite{ShantanuDas2011}
to characterize the zigzag phenomenon that unfolding in traffic control
system. For each pedestrian, continuous but not differential trajectory
is also very common due to interactions with its neighbors as shown
in Figure \ref{zigzag}. Another fact that have been neglected in
lots of previous research is that memory of human beings has been
seldom considered. This is another proof that $d^{\alpha}x/dt^{\alpha}$
is one much better choice than $dx/dt$ in characterizing the movement
of each pedestrian.

Dynamic model of integer order that brought out by Dirk Helbing in
\cite{HelbingDirk1995,Helbing2000} has been extended to the following
dynamic model of fractional order for each pedestrian
\begin{equation}
\left\{ \begin{alignedat}{1}\frac{d^{\alpha}x_{i}}{dt^{\alpha}} & =v_{i},\\
m_{i}\frac{d^{\alpha}v_{i}}{dt^{\alpha}} & =f_{i}^{S}+\sum_{j=1}^{n}f_{ij}^{N}+\sum f_{k}^{W}.
\end{alignedat}
\right.\label{eq:micro-fraction}
\end{equation}
where $x_{i}$ and $v_{i}$ are position and velocity of each pedestrian
(\ref{eq:micro-integer}) respectively, $f_{i}^{S}$ is the self-driven
force towards some desired velocity, $f_{ij}^{N}$ is the interaction
between agent $i$ and its neighbor $j$ and $f_{k}^{W}$ represents
the interactions with environment such as walls or corridors.

\subsection{Fractional Macroscopic Model}

As fractal time and space have been neglected in previous modeling
of crowds, only macroscopic models of integer order have been been
obtained in previous research. Some statistical phenomenons observed
in recent years have forced people to reconsider the effectiveness
of obtained dynamic model of integer order.
\begin{itemize}
\item Distribution of inter-event time that dominating or affecting movement
of single pedestrian can be better approximated by power law rather
than exponential distribution \cite{BruceJ.West2014}. Thus dynamic
models of integer order where exponential distribution has been assumed
are no longer effective any more when confronted with the distribution
of power law. As the hidden dynamics behind distribution of power
law is fractional order, it is much preferred to model crowds of pedestrians
using calculus of fractional order;
\item Different to particles of gases or fluids, pedestrians do not fill
the 2D or 3D space and distribution of the pedestrians is not uniform
in the entire space. Thus space of integer order is not enough to
describe the distribution of pedestrians and fractal space of fractional
order should be included in modeling of crowds of pedestrians.
\end{itemize}
Based on \cite{Kachroo2009} where modeling traffic system has been
considered using calculus of integer order, we try to model crowds
of pedestrians using calculus of fractional order in the followings.

\begin{center}
\begin{figure}[!htb]
\begin{centering}
\includegraphics[scale=0.8]{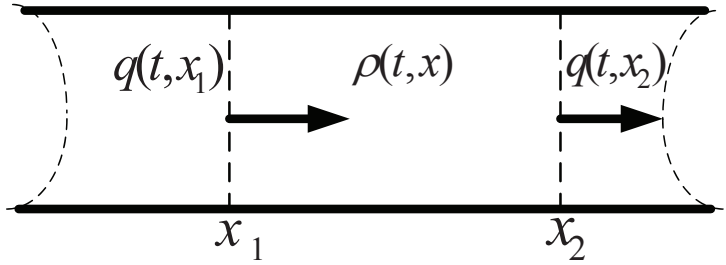}\protect\caption{Conservation of Mass}

\par\end{centering}

\label{traffic}
\end{figure}

\par\end{center}

Denote $\rho(x,t)$ as the density of crowds as shown in Figure \ref{traffic},
then mass of pedestrians between $x=x_{1}$ to $x=x_{2}$ at time
$t$ can be computed as
\begin{equation}
\mbox{mass in}[x_{1},x_{2}]\mbox{ at time}\, t\,:=\int_{x_{1}}^{x_{2}}\rho(x,t)dx^{\beta}.
\end{equation}
For $t\in[t_{1},t_{2}],$ the total mass that enters this domain from
the left boundary at $x=x_{1}$ is given by
\begin{equation}
\mbox{inflow at}\, x_{1}\mbox{from}\, t_{1}\,\mbox{to}\, t_{2}\,:=\!\!\int_{t_{1}}^{t_{2}}\rho(x_{1},t)v(t,x_{1})dt^{\alpha}
\end{equation}
Similarly, the total mass that leaves this domain from the right boundary
at $x=x_{2}$ for $t\in[t_{1},t_{2}]$ is given by
\begin{equation}
\mbox{outflow at}\, x_{2}\,\mbox{from}\, t_{1}\,\mbox{to}\, t_{2}\,:=\!\!\!\int_{t_{1}}^{t_{2}}\!\!\!\rho(x_{2},t)v(t,x_{2})dt^{\alpha}
\end{equation}

As the number of people in the area between $x_{1}$ and $x_{2}$
can change in time due to people crossing the boundary of $x_{1}$
and $x_{2}$. Assuming no pedestrians are created or destroyed, then
the change of number of pedestrians is only due to changes at these
two boundaries. Thus changes of mass of pedestrians in space $[x_{1},x_{2}]$
on time interval $[t_{1},t_{2}]$ is equal to the mass that entering
at $x_{1}$ minus that exiting from $x_{2}$. This conservation can
be describe using
\begin{align*}
\int_{x_{1}}^{x_{2}}\!\!\!\!\rho(t_{2},x)dx^{\beta} & \!\!-\!\!\int_{x_{1}}^{x_{2}}\!\!\!\!\rho(t_{1},x)dx^{\beta}=\int_{t_{1}}^{t_{2}}\!\!\!\!\rho(x_{1},t)v(t,x_{1})dt^{\alpha}\!\!-\!\!\int_{t_{1}}^{t_{2}}\!\!\!\!\rho(x_{2},t)v(t,x_{2})dt^{\alpha}
\end{align*}

The above equation can also be written as the following double integral
form
\begin{equation}
\int_{x_{1}}^{x_{2}}\int_{t_{1}}^{t_{2}}\frac{\partial}{\partial t^{\alpha}}\rho(t,x)+\frac{\partial}{\partial x^{\beta}}[\rho(t,x)v(t,x)]dt^{\alpha}dx^{\beta}=0\label{double}
\end{equation}
Since equation (\ref{double}) should be satisfied for any $t$ and
any $x$, the following fractional order model for crowds of pedestrians
in one dimensional space
\begin{equation}
\frac{\partial}{\partial t^{\text{\ensuremath{\alpha}}}}\rho(t,x)+\frac{\partial}{\partial x^{\beta}}[\rho(t,x)v(t,x)]=0\label{macro model}
\end{equation}
can be derived where fractal time and fractal space have been included
in (\ref{macro model}).
\begin{rem}
Part of the results of fractional model in macro scale has been firstly
brought out in \cite{Ke-CaiCao2012} and are listed here to guarantee
the completeness.
\end{rem}

\begin{rem}
Similar results are also obtained in \cite{Long-FeiWang2014} where
fractional model for traffic flow has been derived using fractional
conservation law. Different to the work of \cite{Long-FeiWang2014}
where dimension of time is $\alpha$, dimension of surface is $2\alpha$
and dimension of volume is $3\alpha$, there are no such restrictions
in our fractional model (\ref{macro model}).
\end{rem}

\subsection{Fractional Micro-Macro Model}

\begin{center}
\begin{figure}[tbh]
\begin{centering}
\includegraphics[scale=0.6]{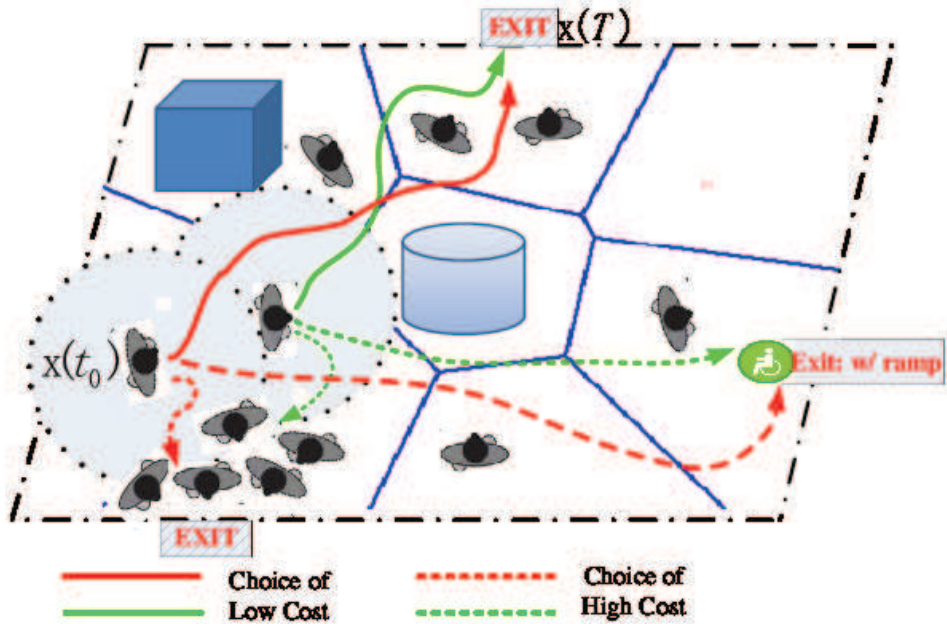}\protect\caption{Movement of pedestrians based on fractional Mean Field Games}

\par\end{centering}

\label{MFG graph}
\end{figure}

\par\end{center}

\subsubsection{Fractional Hamilton-Jacobi-Bellman Equation\label{sub:FHJB}}

For each pedestrian $i$, we assume the following cost function to
be minimized in his movement between initial starting point $x(t_{0})=x_{0}$
and desired location $x(T)$ as shown in Figure \ref{MFG graph}
\begin{equation}
J(t_{0},x_{0})=\underset{v(\cdot)}{inf}\int_{t_{0}}^{T}f(t,x(t),v(t))dt^{\alpha}+h(T,x(T)),\label{eq:HJB cost}
\end{equation}
where convex function $h(T,x(T))$ is the terminal cost, convex even
function $f(t,x(t),v(t))$ describes some different kinds of running
cost between the initial point and destination.
\begin{rem}
A typical quadratic cost function that independent on position of
pedestrians can be selected as $\frac{1}{2}|v|^{2}$ to penalize pedestrians
that moving too fast; Much more generalized running cost functions
that depending on time, position and velocity have been are adopted
in the following derivation of fractional Hamilton-Jacobi-Bellman
Equation.
\end{rem}
Similar to the derivation of Hamilton-Jacobi-Bellman equation of integer
order in optimal control, the fractional Hamilton-Jacobi-Bellman Equation
will be discussed firstly and then optimal velocity will prescribed
for each pedestrian at each time step. Suppose after an infinitesimal
time interval $dt^{\alpha}$, the pedestrian will arrived at one new
place $x_{0}+vdt^{\alpha}$ and thus incurring a travel cost of $f(v)dt^{\alpha}$
where new cost function for the remaining journey described by $J(t_{0}+dt^{\alpha},x_{0}+vdt^{\alpha})$.
The above analysis leads to the following relationship between $J(t_{0},x_{0})$
and $J(t_{0}+dt^{\alpha},x_{0}+vdt^{\alpha})$
\begin{equation}
J(t_{0},x_{0})=J(t_{0}+dt^{\alpha},x_{0}+vdt^{\alpha})+f(v)dt^{\alpha}.\label{eq:heuristic formula}
\end{equation}

Based on Taylor expansion, equation (\ref{eq:heuristic formula})
can be rewritten as
\begin{equation}
\begin{aligned}J(t_{0},x_{0})=J(t_{0},x_{0}) & +dt^{\alpha}[\frac{\partial^{\alpha}}{\partial t^{\alpha}}J(t_{0},x_{0})+v\cdot\frac{\partial^{\beta}}{\partial x^{\beta}}J(t_{0},x_{0})+f(v)],\end{aligned}
\label{eq:Taylor}
\end{equation}
and the optimal problem (\ref{eq:HJB cost}) is now transformed into
finding proper $v$ to minimize
\[
v\cdot\frac{\partial^{\beta}}{\partial x^{\beta}}J(t_{0},x_{0})+f(v).
\]

Considering the fact that $f(\cdot)$ is an even function, the above
minimizing problem is equivalent to the maximizing problem of
\begin{equation}
v\cdot\frac{\partial^{\beta}}{\partial x^{\beta}}J(t_{0},x_{0})-f(v).\label{eq:maxmum problem}
\end{equation}
Based on the Legendre transformation $H:R^{d}\rightarrow R$ of $f:R^{d}\rightarrow R$
by
\begin{equation}
H(p)\text{:=}\underset{v(\cdot)}{sup}\, v\cdot p-f(v)\label{eq:Legendre}
\end{equation}
whose maximum value are functions of $p$. For the maximum problem
of (\ref{eq:maxmum problem}) , we can see that the maximum value
is obtained as $H(\frac{\partial^{\beta}}{\partial x^{\beta}}J(t_{0},x_{0})$)
for some $v$. Then substituting the minimum value $-H(\frac{\partial^{\beta}}{\partial x^{\beta}}J(t_{0},x_{0}))$
into equation (\ref{eq:Taylor}), the following equation
\[
J(t_{0},x_{0})=J(t_{0},x_{0})+dt^{\alpha}[\frac{\partial^{\alpha}}{\partial t^{\alpha}}J(t_{0},x_{0})-H(\frac{\partial^{\beta}}{\partial x^{\beta}}J(t_{0},x_{0}))]
\]
will be satisfied for any $t_{0}$ and any $x_{0}.$ Then the fractional
Hamilton-Jacobi-Bellman Equation is derived as
\begin{equation}
-\frac{\partial^{\alpha}}{\partial t^{\alpha}}J(t_{0},x_{0})+H(\frac{\partial^{\beta}}{\partial x^{\beta}}J(t_{0},x_{0}))=0.\label{eq:fraction HJB}
\end{equation}

From the above discussions, we know that there are some $v$ that
minimize the following expression
\[
v\cdot\frac{\partial^{\beta}}{\partial x^{\beta}}J(t_{0},x_{0})+f(v),
\]
and $\widetilde{v}=-v$ maximize the following expression
\[
v\cdot\frac{\partial^{\beta}}{\partial x^{\beta}}J(t_{0},x_{0})-f(v).
\]
As seen from (\ref{eq:Legendre}), $\widetilde{v}$ as a function
of $p$ should satisfy that

\[
\frac{\partial}{\partial\widetilde{v}}(\widetilde{v}\cdot p-f(\widetilde{v}))=0.
\]

On the other hand, the derivative of $H(p)$ can be obtained as follows
\[
\frac{d}{dp}H(p)=\frac{\partial H}{\partial\widetilde{v}}\frac{\partial\widetilde{v}}{\partial p}+\frac{\partial H}{\partial p}=\widetilde{v},
\]
using chain rule and then the velocity for each pedestrian to move
in the next step is derived as
\[
v=-H^{\prime}(\frac{\partial^{\beta}}{\partial x^{\beta}}J(t_{0},x_{0})).
\]

\subsubsection{Fractional Macro model based on Fractional Mean Field Games}

Based on inspiration of \cite{GeoffroyChevalier2015} on traffic system,
we assume the following utility function for the $i^{th}$ pedestrian
\[
f_{i}^{N}(x_{i},v_{i})=v_{i}(1-F(\frac{1}{N}\sum\omega(x_{j}-x_{i}))),
\]
where the first term $v_{i}$ means that the $i^{th}$ pedestrian
try to arrive his destination as fast as possible; the second term
means that the $i^{th}$ pedestrian adapts his velocity according
to pedestrians around him. Bounded non-negative anticipating function
$\omega(\cdot)$ has been introduced to weight different impacts of
pedestrians in the neighborhood of the $i^{th}$ pedestrian according
to their distances. Thus for the $i^{th}$ pedestrian, cooperative
and competitive interacting with other pedestrians are manifested
through choosing velocity on the next step.

First, we show that the following expression is satisfied
\[
\lim_{N\rightarrow\infty}\frac{1}{N}\sum\omega(x_{j}-x_{i})\rightarrow\int_{0}^{\infty}\rho_{t}(y)\omega(y-x)dy^{\beta},
\]
where $N$ is the number of interacting pedestrians, $\rho_{t}(y)$
is the number of pedestrians in interval $[x,x+dx^{\beta}]$ and $\omega(\cdot)$
is the anticipating function mentioned above.

Denote $\varGamma_{t}^{N}(x)=\frac{1}{N}\sum1_{\{x_{j}<x\}}$ as the
empirical distribution function for the crowds composed of $N$ pedestrians.
Then based on the Lebesgue-Stieltjes integral it can be concluded
that
\[
\frac{1}{N}\sum\omega(x_{j}-x_{i})=\int_{0}^{\infty}\omega(y-x)d\varGamma_{t}^{N}(y).
\]
If there is one non-decreasing right-continuous function $\varGamma_{t}(x)$
such that the following expression is satisfied
\[
\int_{0}^{\infty}\omega(y-x)d\varGamma_{t}^{N}(y)\rightarrow\int_{0}^{\infty}\omega(y-x)d\varGamma_{t}(y)(N\rightarrow\infty).
\]
Then
\[
\frac{1}{N}\sum\omega(x_{j}-x_{i})\rightarrow\int_{0}^{\infty}\rho_{t}(y)\omega(y-x)dy^{\beta}(N\rightarrow\infty)
\]
will be satisfied. As $\rho_{t}(y)$ is the number of pedestrians
in interval $[x,x+dx^{\beta}]$, existence of non-decreasing right-continuous
function $\varGamma_{t}(x)$ can be guaranteed from $d\varGamma_{t}(x)=\rho_{t}(x)dx^{\beta}.$
Thus we can impose the following mean filed payoff function
\[
\begin{aligned}J(t_{0},x_{0},\rho_{t}(x))= & \underset{v(\cdot)}{sup}\int_{t_{0}}^{T}\!\! v(1\!-\! F(\int_{0}^{\infty}\!\!\rho_{t}(y)\omega(y-x)dy^{\beta}))dt^{\alpha}+h(T,x(T))\end{aligned}
\]
for pedestrians that competitively and cooperatively interacting with
other pedestrians.

Based on similar derivations shown in Section \ref{sub:FHJB}, the
following fractional Hamilton-Jacobi-Bellman Equation
\[
-\frac{\partial^{\alpha}}{\partial t^{\alpha}}J(t_{0},x_{0})+H(\frac{\partial^{\beta}}{\partial x^{\beta}}J(t_{0},x_{0},\rho_{t}(x)))=0
\]
can also be obtained for modeling cooperative and competitive crowds
using mean field game theory when the number of pedestrians goes to
infinity.
\begin{rem}
Difference to previous work are listed as followings: \end{rem}
\begin{itemize}
\item Only function of Dirac type and exponential type for $\omega(x_{j}-x_{i})$
have been considered in \cite{GeoffroyChevalier2015}. Anticipating
function of inverse power form
\[
f_{i}^{N}(x_{i},v_{i})=v_{i}[1-F(\frac{1}{N}\sum(\left|x_{j}-x_{i}\right|+1){}^{-2})]
\]
can be included in this paper considering the long range effects in
interacting of multiple pedestrians, where $\frac{1}{N}$ has been
introduced to bound effects of other pedestrians on the $i^{th}$
pedestrian.

\begin{itemize}
\item Mean field games theory is also utilized in \cite{Dogbe2010} for
modeling crowds of pedestrians. But obtained results of \cite{Dogbe2010}
are only restricted to the framework of calculus of integer order
and many statistical characters are not considered such as power law
in distribution of crowds, power law in distribution of inter-event
time and long range interactions among pedestrians.
\end{itemize}
\end{itemize}

\subsubsection{Fractional micro-macro model}

As shown in Figure \ref{Fig:micro-macro model}, the fractional micro-macro
model for crowds of pedestrians using fractional mean field games
can be described as the following backward-forward PDE systems

\begin{equation}
\left\{ \begin{aligned}-\frac{\partial^{\alpha}}{\partial t^{\alpha}}J(t_{0},x_{0},\rho_{t}(x))\!+\! H(\frac{\partial^{\beta}}{\partial x^{\beta}}J(t_{0},x_{0},\rho_{t}(x))) & \!=0,\\
\frac{\partial}{\partial t^{\text{\ensuremath{\alpha}}}}\rho(t,x)\!+\!\frac{\partial}{\partial x^{\beta}}[\rho(t,x)v(t,x)] & \!=0,
\end{aligned}
\right.\label{eq:FMFG}
\end{equation}
and

\begin{equation}
\left\{ \begin{alignedat}{1}\frac{d^{\alpha}x_{i}}{dt^{\alpha}} & =v_{i},\\
m_{i}\frac{d^{\alpha}v_{i}}{dt^{\alpha}} & =f_{i}^{S}+\sum_{j=1}^{n}f_{ij}^{N}+\sum f_{k}^{W}.
\end{alignedat}
\right.\label{eq:micro-fraction-1}
\end{equation}

\begin{center}
\begin{figure}[tbh]
\begin{centering}
\includegraphics[scale=0.85]{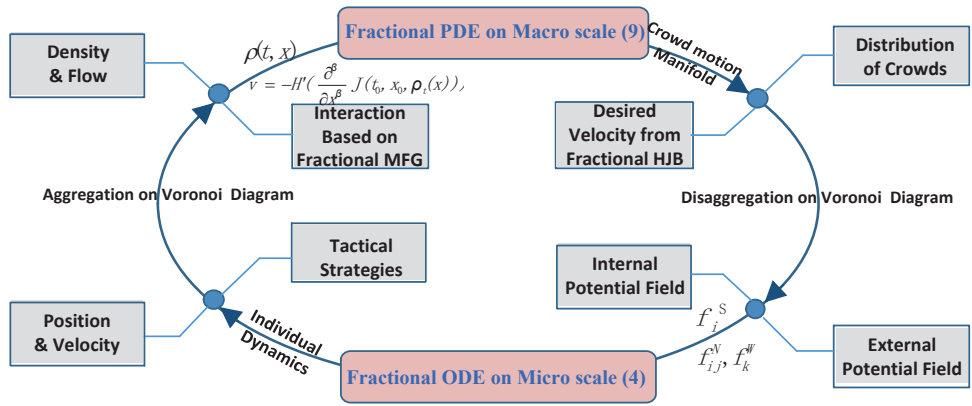}
\par\end{centering}

\protect\caption{Fractional Micro-Macro model of Crowds of Pedestrians}

\label{Fig:micro-macro model}
\end{figure}

\par\end{center}

The fractional microscopic model and fractional macroscopic model
are connected through aggregation and disaggregation on Voronoi Diagram.
From Figure \ref{Fig:micro-macro model}, the followings can be observed.
\begin{enumerate}
\item Movements of each microscopic model are determined by not only internal
potential fields such as the self-driven force towards some desired
velocity described using $f_{i}^{S}$ in (\ref{eq:micro-fraction-1})
but also external interactions from neighbors and environments which
described using $f_{ij}^{N}$ and $f_{k}^{W}$. Some other elements
such as deviations from optimal movement of the whole crowds are also
playing an important role in the movement of each individual pedestrian.
All these information should generated from the dynamic model in macro
scale;
\item Density and velocity that needed in macroscopic model are derived
from aggregation of individual's position and velocity. When the number
of pedestrians goes to infinity, the crowds of pedestrians are treated
as some intelligent flows that described with the help of fractional
MFG as shown in (\ref{eq:FMFG}). For the backward part, $v$ can
be solved from the first line of equation (\ref{eq:FMFG}) under initial
condition on $J(T,X_{T})$ and initial distribution of $\rho_{0}(x)$
derived from aggregation of microscopic model (\ref{eq:micro-fraction-1});
Then substitute the obtained $v$
\[
v=-H^{\prime}(\frac{\partial^{\beta}}{\partial x^{\beta}}J(t_{0},x_{0},\rho_{t}(x)))
\]
into the forward part and $\rho(t,x)$ will be obtained from the second
line of equation (\ref{eq:FMFG}) under initial condition $\rho_{0}(x)$.
\end{enumerate}
Due to the complexity of crowds of pedestrians, fractional microscopic
model and fractional macroscopic model that interacted with each other
have been constructed in this paper. Fractional mean field games have
also been utilized in describing the macroscopic model when the number
of pedestrians goes to infinity.
\begin{rem}
To the author's knowledge, the paper is one of the first works applying
fractional mean field games to fractional macroscopic and microscopic
model for competitive and cooperative crowds of pedestrians. Although
some theoretical work has been obtained, a lot of work are waiting
for further efforts such as existence and uniqueness of solution,
rate of convergence and stability of desired equilibrium.
\end{rem}

\section{Simulation Results\label{sec:Simulation-Results}}

Considering unexpected or dangerous events in real-life experiment,
only some initial simulation results are conducted to show the differences
between model of fractional order and model of integer order in macro
scale and micro scale. Due to the difficulties caused when the number
of pedestrians goes to infinity, simulation results in macro scale
and micro scale are separated in the following subsections. All we
want to show is that calculus of fractional order has offered us much
more freedom in describing complex phenomenon or dynamics such as
crowds of pedestrians. It is much preferred to choose different model
according to different scenarios and there are a lot of interesting
problems needing to be considered in future.

\subsection{Fractional Macroscopic Model}

\subsubsection{\label{sub:Simulation-in-closed}Simulation in closed and square
area without exit}

Simulation results on fractional macroscopic model (\ref{macro model})
are firstly conducted where $\beta=1$ are imposed for simplicity.
Lax-Friedrichs Scheme has been used to approximate the spatial derivatives
in solving the nonlinear partial differential equations due to its
efficiency in computation. Based on Lax-Friedrichs Scheme, the following
PDE on 2D plane
\[
\begin{aligned}\frac{\partial}{\partial t^{\text{\ensuremath{\alpha}}}}\rho(t,x,y) & +\frac{\partial}{\partial x}[\rho(t,x,y)v(t,x,y)]+\frac{\partial}{\partial y}[\rho(t,x,y)v(t,x,y)]=0\end{aligned}
\]
has been transformed into

\[
\begin{array}{c}
\frac{\partial}{\partial t^{\text{\ensuremath{\alpha}}}}\rho(t,x,y)+\frac{1}{2Dx}[\rho(t,x+1,y)v(t,x+1,y)\\
-\rho(t,x-1,y)v(t,x-1,y)]+\frac{1}{2Dy}[\rho(t,x,y+1)v(t,x,y+1)-\rho(t,x,y-1)v(t,x,y-1)]=0
\end{array}
\]
in the simulations.

Under the following initial Gaussian distribution
\[
\rho(x,y,0)=C\exp(-(x-a)^{2}-(y-b)^{2}),
\]
where $C=1$ is the density value and $(a,b)$ determines the center
of initial density distribution. Average speed of free flow has been
chosen to be $v_{x}=v_{y}=1.36ms^{-1}$as done in many previous studies
for pedestrians. Pedestrians have also been assumed to move freely
within a square area with no obstacles and no exits in the first simulations.

Simulation results for $\alpha=0.6$ and $\alpha=1$ are shown in
Figure \ref{macro mesh graph} to \ref{macro contour graph} and Figure
\ref{macro mesh graph-integer} to \ref{macro contour graph-integer},
respectively. From Figure \ref{macro mesh graph} and Figure \ref{macro mesh graph-integer},
it can be concluded that pedestrians described using fractional model
are much scattered in the closed square area than that described using
model of integer order. Same conclusions can also be obtained from
comparisons between Figure \ref{macro contour graph} and Figure \ref{macro contour graph-integer}.
Other fractional orders can also be tested using the methods proposed
in this paper but data from reality are much preferred to find the
proper orders for modeling the crowds of pedestrians in macro scale.

\begin{center}
\begin{figure}[tp]
\noindent \begin{centering}
\includegraphics[scale=0.8]{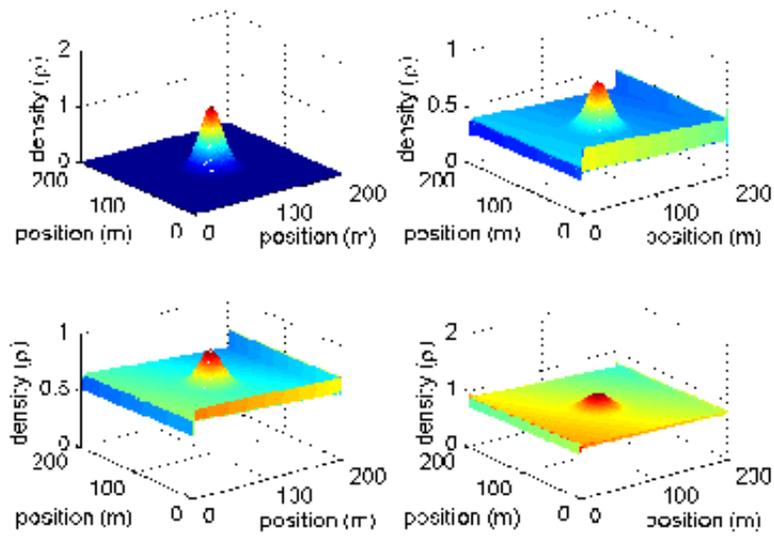}
\par\end{centering}

\protect\caption{\selectlanguage{english}%
Density response for crowds of pedestrians with $\alpha=0.6$ using
Lax-Friedrichs Scheme\selectlanguage{american}%
}

\label{macro mesh graph}
\end{figure}

\par\end{center}

\begin{center}
\begin{figure}[tp]
\noindent \begin{centering}
\includegraphics[scale=0.8]{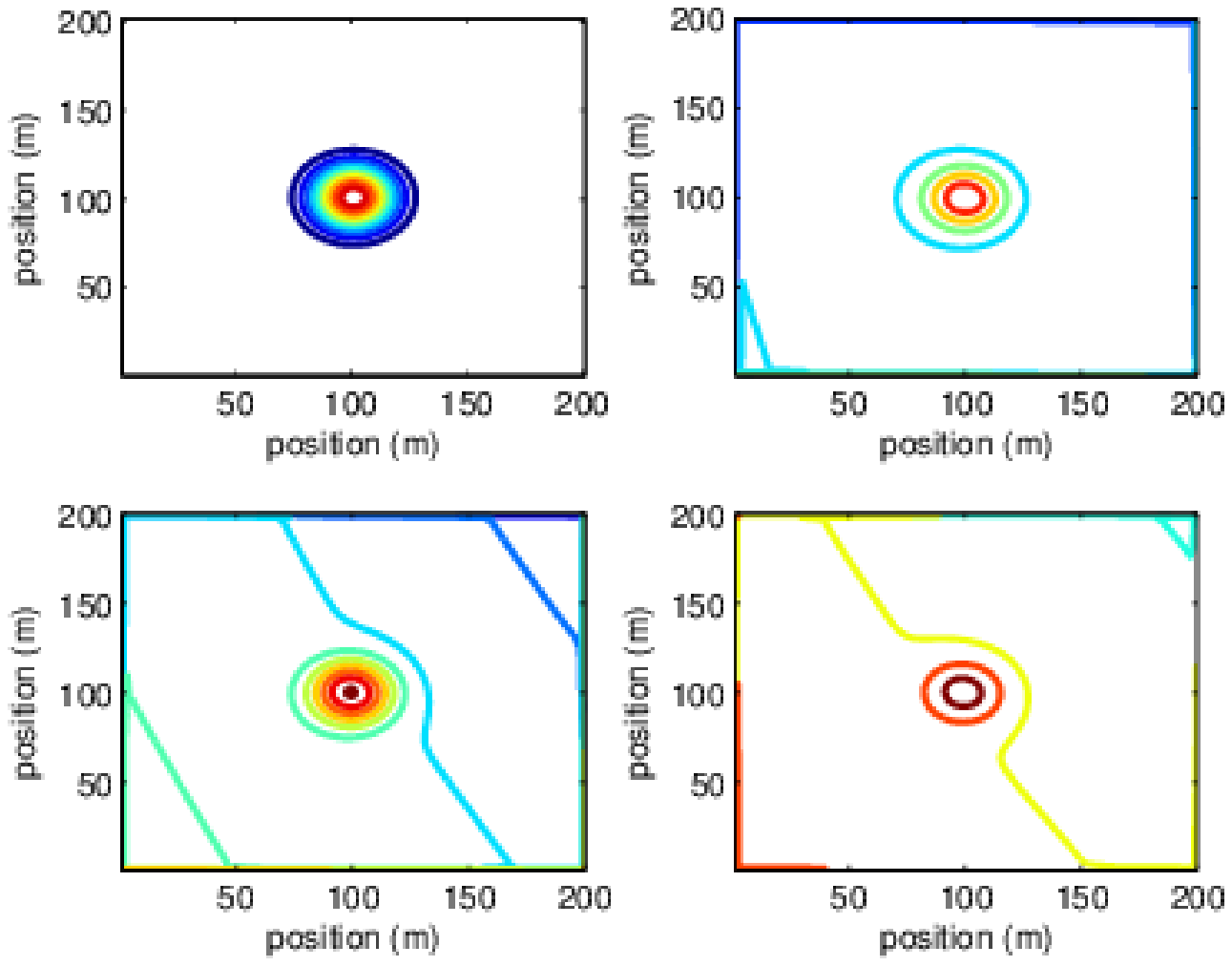}
\par\end{centering}

\protect\caption{\selectlanguage{english}%
Contour of the density response for crowds of pedestrians with $\alpha=0.6$
using Lax-Friedrichs Scheme\label{macro contour graph}\selectlanguage{american}%
}
\end{figure}

\par\end{center}

\begin{center}
\begin{figure}[tp]
\noindent \begin{centering}
\includegraphics[scale=0.8]{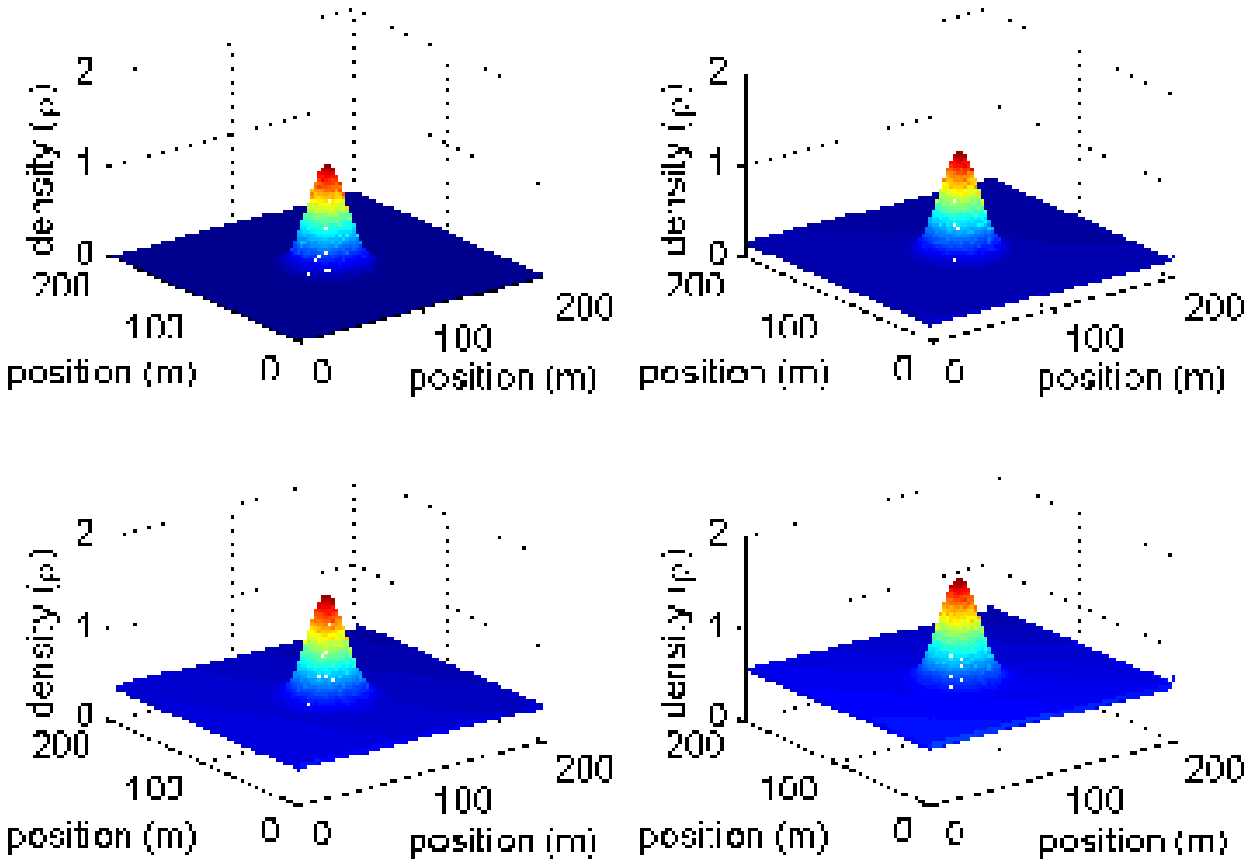}
\par\end{centering}

\protect\caption{\selectlanguage{english}%
Density response for crowds of pedestrians of integer order using
Lax-Friedrichs Scheme\selectlanguage{american}%
}

\label{macro mesh graph-integer}
\end{figure}

\par\end{center}

\begin{center}
\begin{figure}[tp]
\noindent \begin{centering}
\includegraphics[scale=0.8]{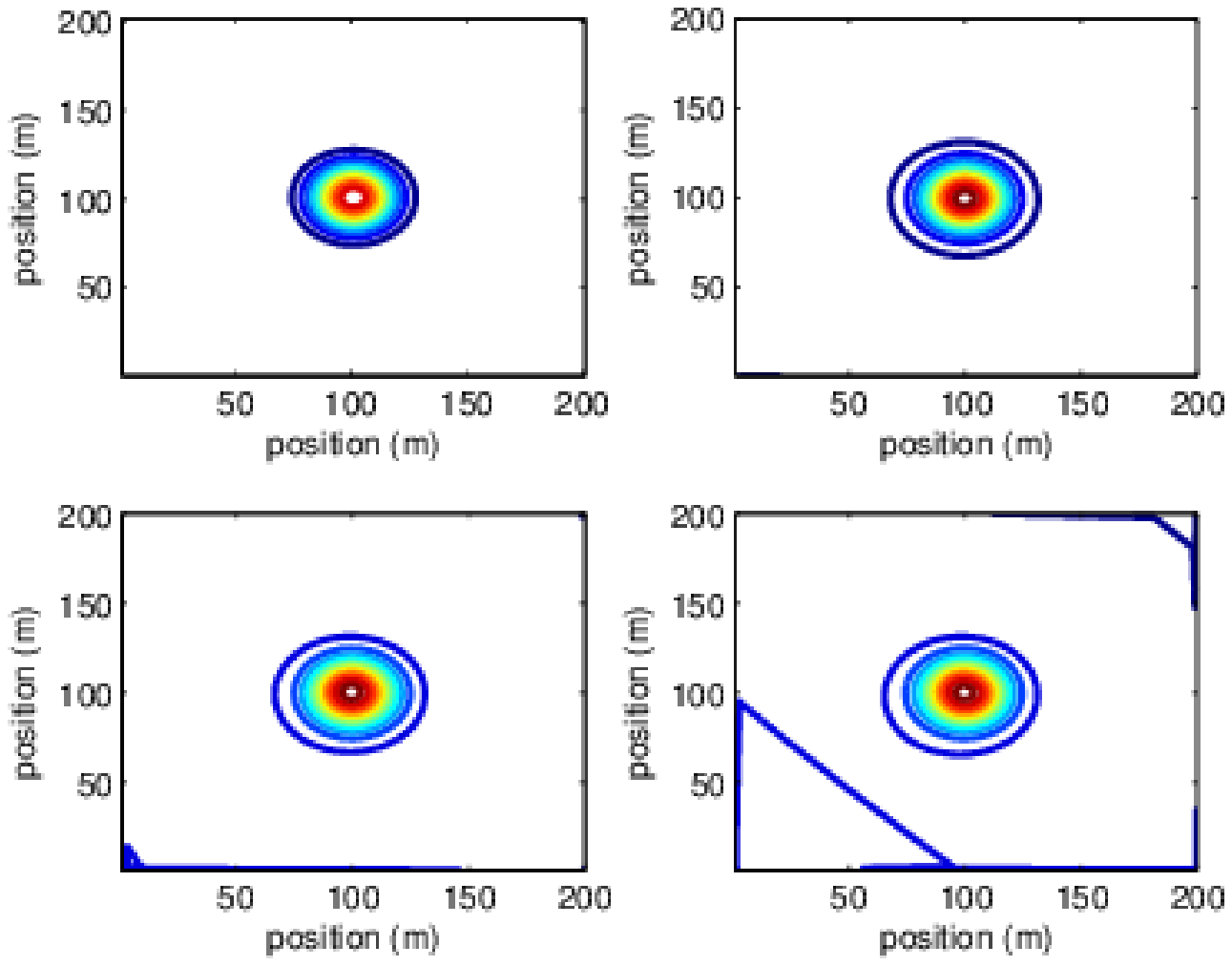}
\par\end{centering}

\protect\caption{\selectlanguage{english}%
Contour of the density response for crowds of pedestrians of integer
order using Lax-Friedrichs Scheme\label{macro contour graph-integer}\selectlanguage{american}%
}
\end{figure}

\par\end{center}

\subsubsection{Simulation in closed and square area with one exit}

Based on results obtained in Section \ref{sub:Simulation-in-closed},
the following dynamic model has been simulated for pedestrians in
closed and square area with one exit

\[
\left\{ \begin{aligned}\frac{\partial}{\partial t^{\text{\ensuremath{\alpha}}}}\rho(t,x,y)+\frac{\partial}{\partial x} & [\rho(t,x,y)v(t,x,y)]\\
+\frac{\partial}{\partial y} & [\rho(t,x,y)v(t,x,y)]=0,\\
v_{t}+vv_{x}\,\,\,\, & =\frac{V-v}{\tau}-\frac{C_{0}^{2}}{\rho}\rho_{x},\\
u_{t}+uu_{y}\,\,\,\, & =\frac{U-u}{\tau}-\frac{C_{0}^{2}}{\rho}\rho_{y},
\end{aligned}
\right.
\]
where $C_{0}=0.8$ is the anticipation term that describes the response
of pedestrians to density of people and $V$ and $U$ are some desired
velocity that obtained for the crowds. In order to lead the crowd
moving toward the exit, the desired velocity $V$ and $U$ are selected
as done in \cite{Kachroo2008b}

\[
\left\{ \begin{array}{cc}
V= & V(\rho)\frac{x_{e}-x_{i}}{\sqrt{(x_{e}-x_{i})^{2}+(y_{e}-y_{i})^{2}}}\\
U= & U(\rho)\frac{y_{e}-y_{i}}{\sqrt{(x_{e}-x_{i})^{2}+(y_{e}-y_{i})^{2}}}
\end{array}\right.
\]
where $V(\rho)$ and $U(\rho)$ are the flux-density relationship
for Greenshield's model.

Simulation results are shown in Figure \ref{evacuation-.9} and Figure
\ref{evacuation-1} where dynamic model with fractional order $0.85$
and $1$ are used. Simulation results show that the density of pedestrians
around the exit is much lower in model of fractional order than that
obtained using model of integer order. In simulations, the authors
found that the stable density of pedestrians is depending on the fractional
order selected in the simulation and how to choose the best order
to model the dynamics of crowds is an interesting problem that is
worthy of further consideration in future research.

\begin{center}
\begin{figure}[tp]
\noindent \begin{centering}
\includegraphics[scale=0.8]{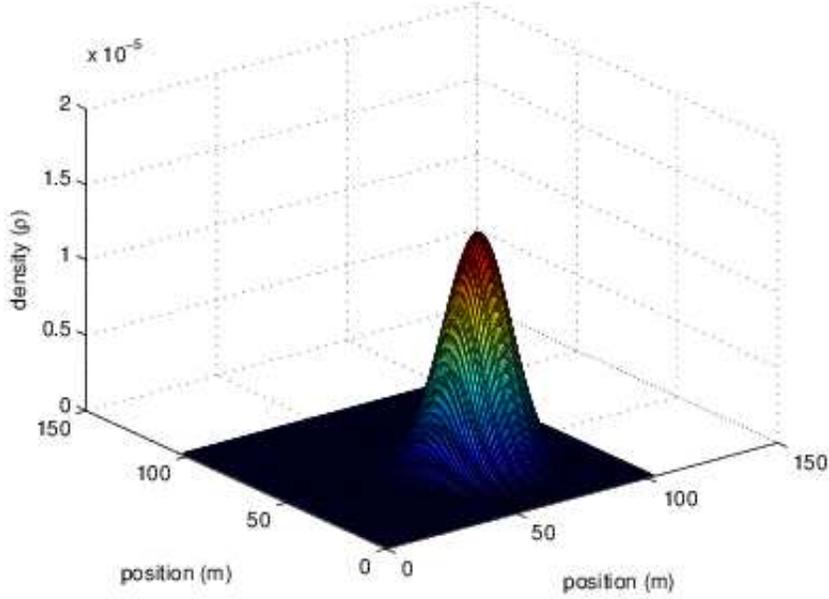}
\par\end{centering}

\protect\caption{\selectlanguage{english}%
Density response for crowds of pedestrians with $\alpha=0.85$ using
Lax-Friedrichs Scheme\selectlanguage{american}%
}

\label{evacuation-.9}
\end{figure}

\par\end{center}

\begin{center}
\begin{figure}[tp]
\noindent \begin{centering}
\includegraphics[scale=0.8]{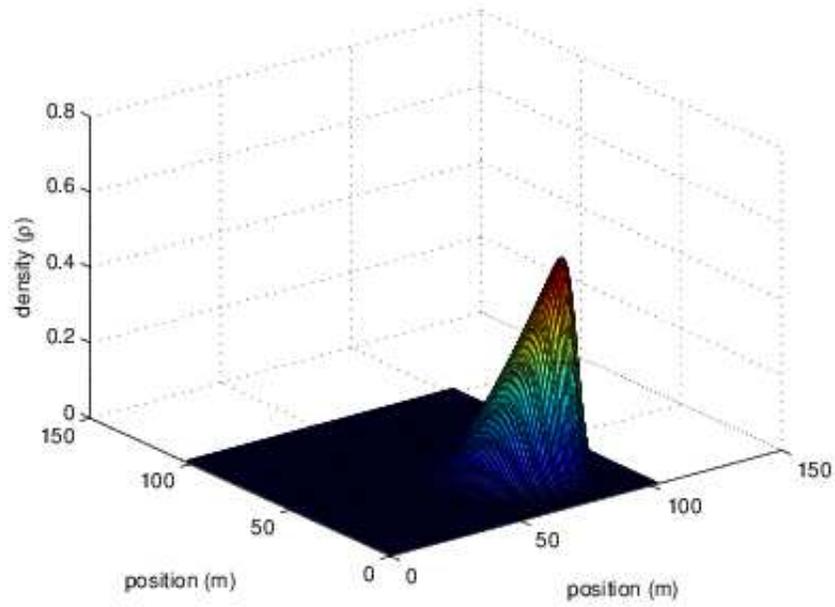}
\par\end{centering}

\protect\caption{\selectlanguage{english}%
Contour of the density response for crowds of pedestrians with $\alpha=1$
using Lax-Friedrichs Scheme\label{evacuation-1}\selectlanguage{american}%
}
\end{figure}

\par\end{center}

\subsection{Fractional Microscopic Model}

In this section, six pedestrians with fractional order $\alpha\in(0,1)$,
$\alpha=1$ and $\alpha\in(1,2)$ are employed respectively to show
their effects on pedestrian's evacuation process. Simulations of crowds
of pedestrians with fractional order $\alpha=0.6$, $\alpha=1$ and
$\alpha=1.3$ are shown in Figure \ref{game graph  point6}, Figure
\ref{game graph 1} and Figure \ref{game graph 1point3} respectively.
Results show that all agents firstly reach consensus through interacting
with their neighbors without games. But parts of them changed their
desired value and fragmentation phenomenon are observed through these
simulations after some penalty terms are injected into the simulations.
Obtained simulation results have shown that pedestrians with different
orders have different performance. Thus Fractional Calculus has provided
us much more freedom in analysis and control of this kind of complex
system. How to quantitatively characterize the relationship between
order of fractional model, fractional controller and fractional games
are interesting topics to be considered for the authors.

\begin{center}
\begin{figure}[tp]
\begin{centering}
\includegraphics[scale=0.5]{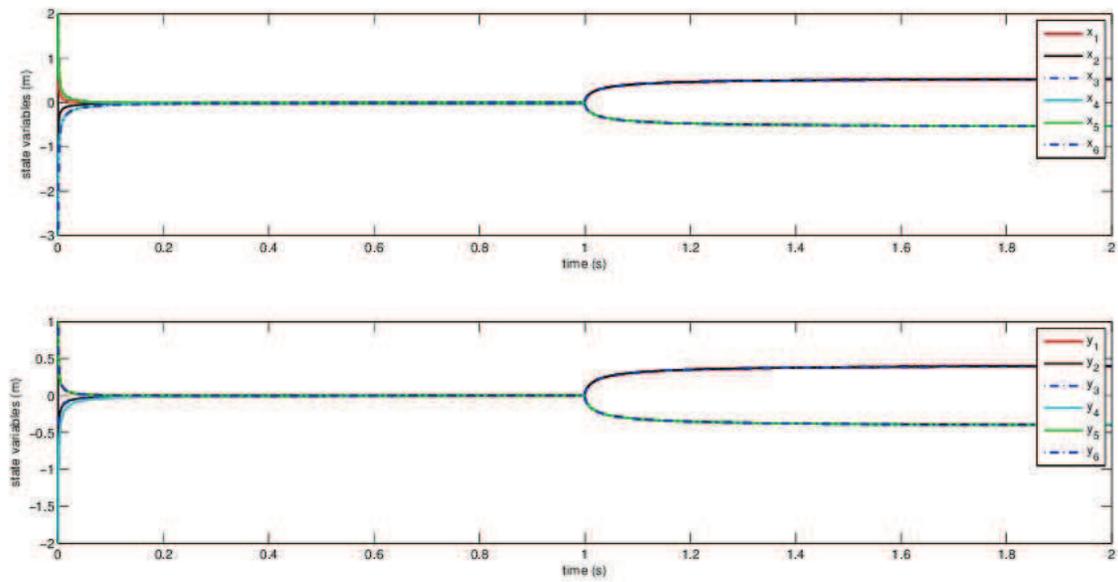}
\par\end{centering}

\protect\caption{\selectlanguage{english}%
Responses of six pedestrians with $\alpha=0.6$\selectlanguage{american}%
}

\label{game graph  point6}
\end{figure}

\par\end{center}

\begin{center}
\begin{figure}[tp]
\begin{centering}
\includegraphics[scale=0.5]{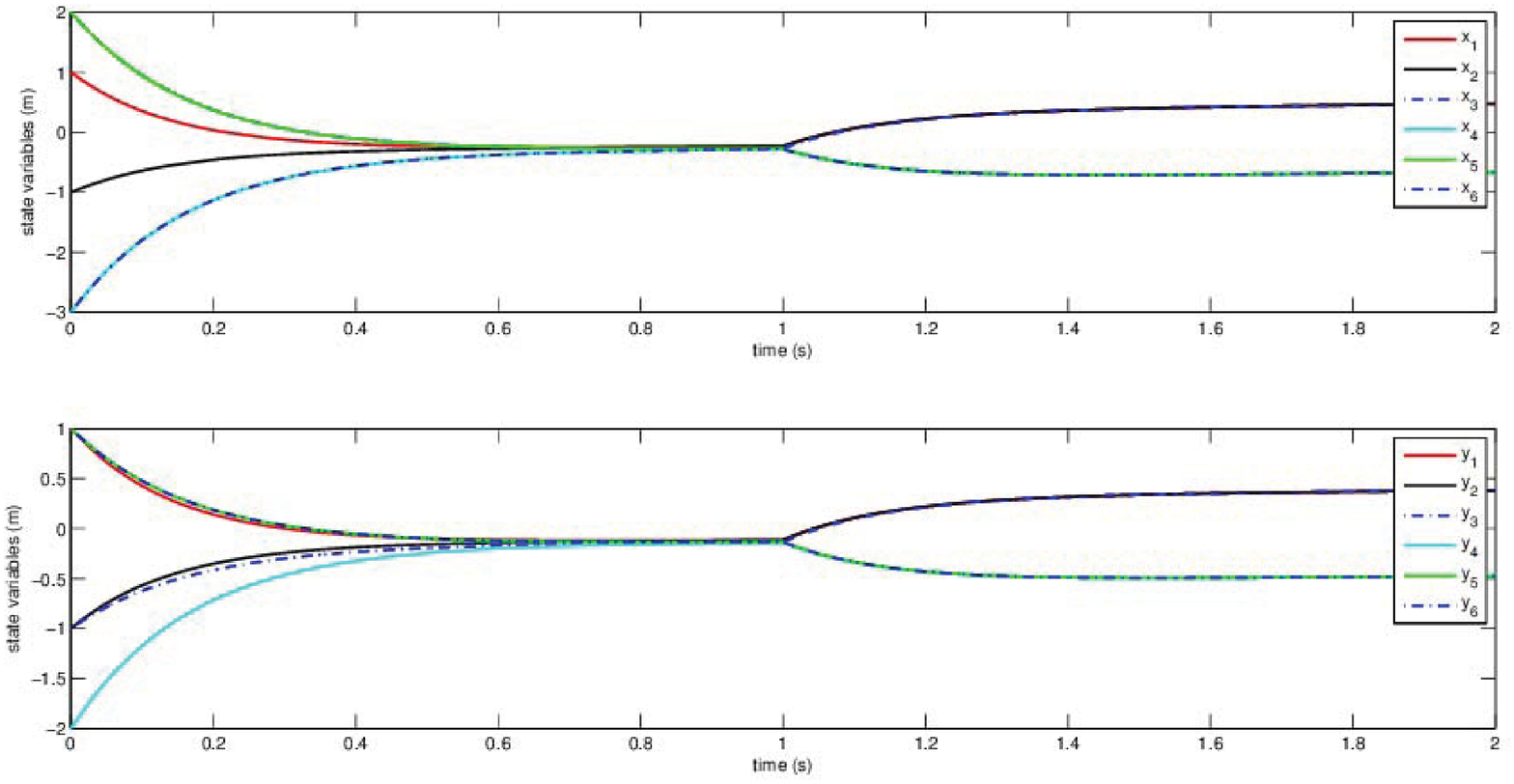}
\par\end{centering}

\protect\caption{\selectlanguage{english}%
Responses of six pedestrians with $\alpha=1$\selectlanguage{american}%
}

\label{game graph 1}
\end{figure}

\par\end{center}

\begin{center}
\begin{figure}[tp]
\begin{centering}
\includegraphics[scale=0.5]{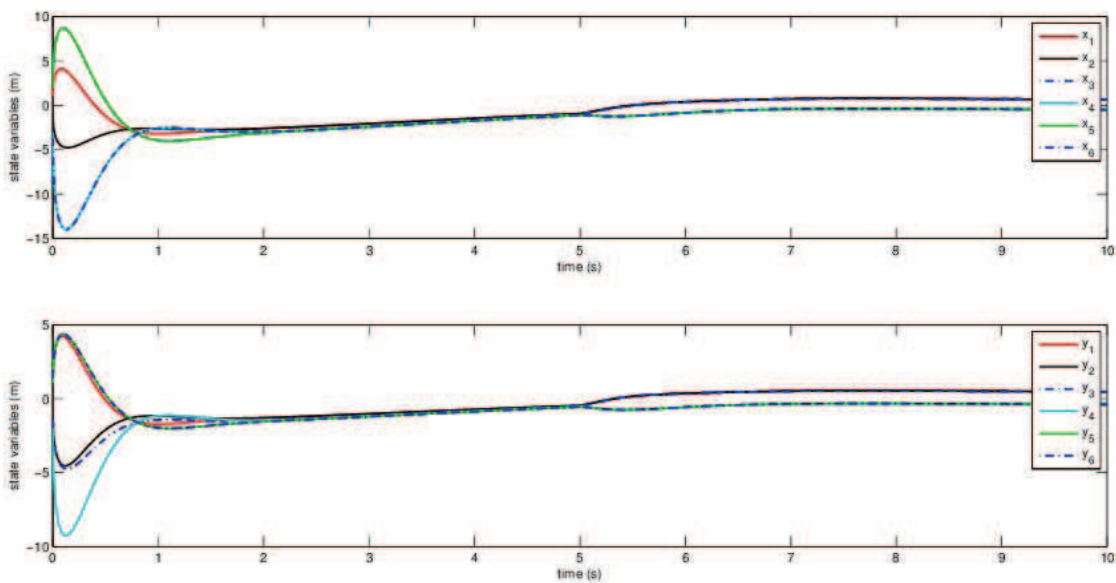}
\par\end{centering}

\protect\caption{\selectlanguage{english}%
Responses of six pedestrians with $\alpha=1.3$\selectlanguage{american}%
}

\label{game graph 1point3}
\end{figure}

\par\end{center}

\section{Conclusions}

Modeling of crowds of pedestrians have been considered in this paper
from the view of Fractional Calculus. Not only fractional microscopic
models but also fractional macroscopic models have been proposed in
this paper. Fractional mean field games theory have been introduced
in the modeling of crowds of pedestrians and coupled PDEs composed
of fractional backward part and fractional forward part have been
investigated. Although some theoretical results and some initial simulations
are presented in this paper, there are much more work unexplored along
this topic, such as solution of fractal MFG systems, stability of
the fractal MFG system and performance of this fractal system, controller
design based on mean field, performance evaluation of dynamic crowds
and security problems related to control of crowds.

\bibliographystyle{unsrt}

\end{document}